\documentclass[runningheads]{llncs}
\usepackage{graphicx}
\usepackage{array}
\newcolumntype{P}[1]{>{\centering\arraybackslash}p{#1}}
\usepackage{amsmath,amssymb}
\usepackage{wrapfig}

\usepackage{color}
\definecolor{mattblue}{RGB}{23,55,180}
\definecolor{mattorange}{RGB}{178,101,24}
\newcommand{\eat}[1]{}
\newcommand{\invivo}{\textit{in vivo} }

\begin{document}
\title{Domain Adaptation for Ultrasound Beamforming}
%
%
\author{Jaime Tierney \and
Adam Luchies \and
Christopher Khan \and
Brett Byram \and
Matthew Berger}


%
\authorrunning{J. Tierney et al.}
%
\institute{Vanderbilt University, Nashville, TN 37235, USA 
\email{jaime.e.tierney@vanderbilt.edu}\\}
\maketitle              

\begin{abstract}

Ultrasound B-Mode images are created from data obtained from each element in the transducer array in a process called beamforming. The beamforming goal is to enhance signals from specified spatial locations, while reducing signal from all other locations. On clinical systems, beamforming is accomplished with the delay-and-sum (DAS) algorithm. DAS is efficient but fails in patients with high noise levels, so various adaptive beamformers have been proposed. Recently, deep learning methods have been developed for this task. With deep learning methods, beamforming is typically framed as a regression problem, where clean, ground-truth data is known, and usually simulated. For \textit{in vivo} data, however, it is extremely difficult to collect ground truth information, and deep networks trained on simulated data underperform when applied to \textit{in vivo} data, due to domain shift between simulated and \textit{in vivo} data. In this work, we show how to correct for domain shift by learning deep network beamformers that leverage both simulated data, and unlabeled \textit{in vivo} data, via a novel domain adaption scheme. A challenge in our scenario is that domain shift exists both for noisy input, and clean output. We address this challenge by extending cycle-consistent generative adversarial networks, where we leverage maps between synthetic simulation and real \textit{in vivo} domains to ensure that the learned beamformers capture the distribution of both noisy and clean \textit{in vivo} data. We obtain consistent \textit{in vivo} image quality improvements compared to existing beamforming techniques, when applying our approach to simulated anechoic cysts and \textit{in vivo} liver data. 

\end{abstract}

\section{Introduction}

Ultrasound imaging is an indispensable tool for clinicians because it is real-time, cost-effective, and portable. However, ultrasound image quality is often suboptimal due to several sources of image degradation that limit clinical utility. Abdominal imaging is particularly challenging because structures of interest are beneath various tissue layers which can corrupt received channel signals due to phenomena like off-axis scattering or reverberation clutter \cite{dahl2014}. 

Several advanced beamforming methods have been proposed to address this problem. Compared to conventional delay-and-sum (DAS) beamforming which applies constant delays and weights to the received channel data for a given spatial location, advanced methods aim to adaptively adjust the received channel data to enhance signals of interest and suppress sources of image degradation. This adaptive beamforming has been accomplished through coherence-based techniques \cite{li2003,lediju2011}, adaptive apodization schemes \cite{synnevag2007,holfort2009}, as well as through model-based approaches \cite{byram2014,byram2015,dei2017,dei2018}. Although effective, most of these advanced methods are computationally intensive and/or limited by user defined parameters. For example, despite achieving consistently superior image quality compared to DAS as well as other advanced techniques, a model-based approach called aperture domain model image reconstruction (ADMIRE) is exceedingly computationally complex, preventing real-time adjustable implementations \cite{dei2019}. 

More recently, there has been growing interest in using deep learning methods for ultrasound beamforming. These efforts generally fall under two categories, the first having the goal of using neural networks to reconstruct fully sampled data from some form of sub-sampled receive channel data \cite{perdios2017,gasse2017,yoon2018,senouf2018,khan2019}. The second has the goal of using neural networks to perform a form of adaptive beamforming using physical ground truth information during training \cite{luchies2018,luchies2019,nair2018,nair2019,hyun2019,zhuang2019}. The former is restricted to the desired fully sampled DAS or advanced beamforming output while the latter is theoretically capable of surpassing DAS or advanced beamformer performance. Other adaptive beamforming deep learning methods have been proposed that use an advanced beamformer as ground truth \cite{simson2019}, which despite providing improvements to DAS, are constrained by the performance of the adaptive beamformer that they mimic. 

One of the main limitations of deep learning for performing adaptive ultrasound beamforming is the lack of ground truth information \textit{in vivo}. Previous efforts have primarily relied on simulations to generate labeled training data sets \cite{luchies2018,nair2018,hyun2019}. Unlike \textit{in vivo} data, which is also costly to obtain, simulations can be controlled to generate unlimited amounts of realistic training data for which ground truth information is known. Network beamformers trained with simulated data have shown some success at generalizing to \textit{in vivo} data \cite{luchies2018,luchies2019,hyun2019}. However, despite sophisticated ultrasound simulation tools, a domain shift still exists between simulated and \textit{in vivo} data, which ultimately limits network beamformer performance.

To address these limitations, we propose a novel domain adaptation scheme to incorporate \textit{in vivo} data during training. Our approach uses cycle-consistent generative adversarial networks (CycleGANs) which learn maps between two data distributions absent of paired data~\cite{zhu2017}.  GANs have been proposed previously in the context of ultrasound beamforming \cite{nair2019} but, to the best of our knowledge, have never been considered for performing domain adaption with real unlabeled \textit{in vivo} data. Further, although CycleGANs have been leveraged to address domain shift in inputs for recognition tasks~\cite{hoffman2017}, in our scenario domain shift exists for \textit{both} inputs and outputs. We mitigate both sources of domain shift by composing CycleGAN maps with domain-specific regressors to effectively learn deep \textit{in vivo} beamformers. \eat{We leverage CycleGANs to generate maps between simulated and \textit{in vivo} ultrasound channel data domains, to learn an \textit{in vivo} beamformer absent of any labeled \textit{in vivo} data.} We develop and evaluate our approach using simulated anechoic cyst and \textit{in vivo} liver data, and compare our approach to conventional DAS, deep neural networks (DNNs) trained using simulated data only, as well as established coherence and model-based advanced beamforming techniques. 

\section{Methods}

\subsection{Domain Adaptation}

The basic intuition behind our approach is to, simultaneously, learn both \emph{regressors} for beamforming, as well as \emph{maps} that allow us to transform simulated channel data into corresponding \invivo data, and vice versa. More concretely, we denote $S$ as our simulated domain, $T$ as our \invivo domain, $x_s$ and $x_t$ refer to input channel data for simulated and \invivo data, respectively, and $y_s$ and $y_t$ refer to output channel data for their respective $S$ and $T$ domains. All channel data are $d$-length signals that we treat as $d$-dimensional vectors. We are provided a set of simulated input and output pairs, each denoted $(x_s,y_s)$, as well as a set of \invivo inputs, but no corresponding outputs, denoted $x_t$. Our main goal is to learn a function $F_t : \mathbb{R}^d \rightarrow \mathbb{R}^d$ that serves as a beamformer for \invivo data. The challenge we address in this work is how to learn $F_t$ absent of any \invivo outputs $y_t$, wherein our goal is to produce $(x_t,y_t)$ pairs that approximate actual input/output \invivo pairs, and can be used to train $F_t$.

There are several ways to address this problem. Previous deep beamformer approaches~\cite{luchies2018} learn $F_t$ from simulated pairs $(x_s,y_s)$. However, applying $F_t$ to \invivo inputs leads to \emph{domain mismatch}, as the data distributions of $x_s$ and $x_t$ differ, due to simulation modeling assumptions that do not always capture \invivo physics. Thus, the starting point for our approach is to mitigate domain shift in the inputs. Specifically, our aim is to learn a mapping $G_{S \rightarrow T}$ that takes a given $x_s$ and maps it into a corresponding \invivo input $x_t$. A common method for learning maps between domains is the use of generative adversarial networks~\cite{goodfellow2014}, specifically for image translation tasks~\cite{isola2017}. Such methods, however, assume paired data, where in our case simulated and \invivo data are not paired. For unpaired data, the CycleGAN approach of Zhu et al.~\cite{zhu2017} proposes to learn maps from $S$ to $T$, $G_{S \rightarrow T}$, \emph{and} from $T$ to $S$, $G_{T \rightarrow S}$, enforcing cycle-consistency between maps. Specifically, for $G_{S \rightarrow T}$ we formulate the adverarial loss as follows:
\begin{equation}
L_{G_{S \rightarrow T}}(G_{S \rightarrow T},D_{T})=\mathbb{E}_{x_{t}\sim X_{T}}[\log D_{T}(x_t)] + \mathbb{E}_{x_{s}\sim X_{S}}[1-\log D_{T}(G_{S \rightarrow T}(x_s))],
\label{generalGANEq}
\end{equation}
\noindent where $D_T$ is the discriminator, tasked with distinguishing real \invivo data from fake \invivo data produced by $G_{S \rightarrow T}$, while $X_S$ and $X_T$ are the distributions for simulated and \invivo data respectively. We may define a similar adversarial loss $L_{G_{T \rightarrow S}}$ for $G_{T \rightarrow S}$, with its own discriminator $D_S$. A cycle consistency regularization is also incorporated to ensure similarity between reconstructed signals and original data~\cite{zhu2017}, defined as
\begin{equation}
\begin{split}
L_{cyc}(G_{S \rightarrow T},G_{T \rightarrow S})=\mathbb{E}_{x_{s}\sim X_{S}}[||G_{T \rightarrow S}(G_{S \rightarrow T}(x_{s}))-x_{s}||_{1}] \\ + 
\mathbb{E}_{x_{t}\sim X_{T}}[||G_{S \rightarrow T}(G_{T \rightarrow S}(x_{t}))-x_{t}||_{1}].
\end{split}
\label{cycLoss}
\end{equation}

\begin{figure}[t]
	\centering
	\includegraphics[width=4in]{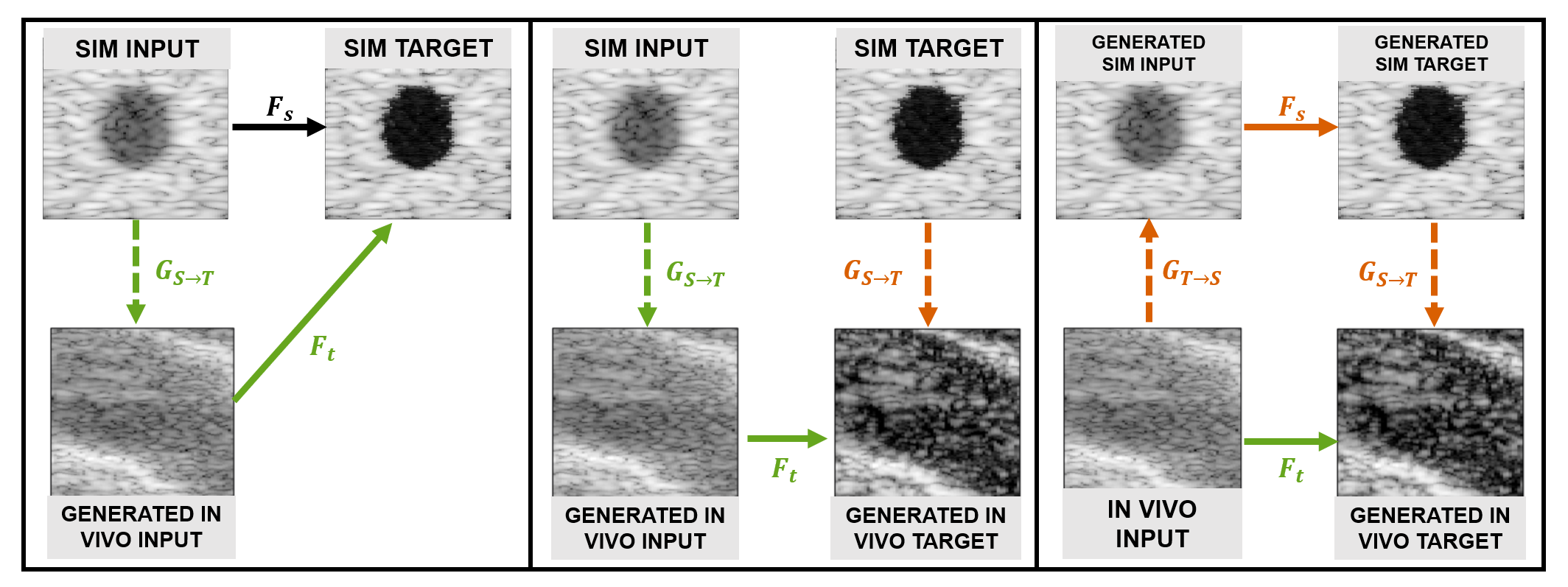}
	\caption{Summary of input and output domain adaptation for training an \textit{in vivo} beamformer, $F_{t}$. Green and orange arrows indicate \textit{in vivo} input and output data generation, respectively. The left schematic summarizes previous efforts for which output domain adaptation was not considered \cite{hoffman2017}. In comparison, data used to compute $L_{F_{T1}}$ and $L_{F_{T2}}$ for the proposed method are indicated by the middle and right schematics, respectively. Data used to compute $L_{F_{S}}$ is indicated by the black arrow on the left.} \label{methodsFigDA}
\end{figure}

The above discriminators and maps can be jointly optimized with $F_t$, where we may provide generated, paired \invivo data via $(G_{S \rightarrow T}(x_s), y_s)$, highlighted in Fig.~\ref{methodsFigDA}(left). This is at the core of the CyCADA method~\cite{hoffman2017}. CyCADA is focused on recognition problems, e.g. classification and semantic segmentation, and thus the target output used (class label) can be easily leveraged from the source domain. However, for our scenario this is problematic, as domain shift still exists between $y_s$ and $y_t$, and thus training on $(G_{S \rightarrow T}(x_s), y_s)$ necessitates $F_t$ to \emph{both} resolve domain gap, and beamform. 

In contrast to CyCADA \cite{hoffman2017}, we would rather have $F_t$ focus \emph{only} on beamforming. To address this, we leverage our domain maps for which we make the assumption that the domain shift in the inputs is identical to domain shift in the outputs, and also introduce a learned function $F_s$ for beamforming simulated data, to arrive at the following \invivo beamforming losses, as illustrated in Fig.~\ref{methodsFigDA}:
\begin{equation}
L_{F_{S}}=\mathbb{E}_{x_{s}\sim X_{S}}[||F_s(x_{s})-y_{s}||_{l}]
\label{simLoss}
\end{equation}
\begin{equation}
L_{F_{T1}}=\mathbb{E}_{x_{s}\sim X_{S}}[||F_t(G_{S \rightarrow T}(x_{s}))-G_{S \rightarrow T}(y_{s})||_{l}],
\end{equation}
\begin{equation}
L_{F_{T2}}=\mathbb{E}_{x_{t}\sim X_{T}}[||F_t(x_{t})-G_{S \rightarrow T}(F_s(G_{T \rightarrow S}(x_{t})))||_{l}].
\end{equation}
\noindent Intuitively, $L_{F_{T1}}$ ensures $F_t$ can beamform generated \invivo data produced from paired simulated data. The term $L_{F_{T2}}$ ensures that $F_t$ can beamform real \invivo data. In Fig.~\ref{methodsFigDA}, example fully reconstructed simulated anechoic cyst and \textit{in vivo} images are used, however our networks operate on aperture domain signals, as depicted in Fig. \ref{methodsFigDatSummary} and described in more detail in the following section.

Our full loss may be formulated as follows:
\begin{equation}
	L=\underbrace{\lambda_{{s}}L_{G_{S \rightarrow T}}+\lambda_{{t}}L_{G_{T \rightarrow S}}+\lambda_{c}L_{cyc}}_\text{GAN}+\underbrace{\lambda_{F_{S}}L_{F_S}+\lambda_{F_{T}}(L_{F_{T1}}+L_{F_{T2}})}_\text{Regressor},
\label{totalLoss}
\end{equation}
\noindent where we simultaneously optimize for discriminators, generators, and regressors. We set GAN-related weights based on Hoffman et al. \cite{hoffman2017} (i.e., $\lambda_s$=2, $\lambda_t$=1,$\lambda_c$=10), while the regressor loss weights were empirically chosen and both set to 1. We also regularize discriminators based on the approach of Mescheder et al.~\cite{mescheder2018training}. Furthermore, in order to ensure that the regressors utilize distinct, and shared, features from the two domains, we learn a single regressor $F : \mathbb{R}^d \times \mathbb{R}^d \times \mathbb{R}^d \rightarrow \mathbb{R}^d$ using the augmentation method of Daum\'e~\cite{daum2007}, such that $F_s(x_s) = F(x_s,x_s,0)$ and $F_t(x_t) = F(x_t,0,x_t)$, e.g. the first argument captures domain-invariant features, while the second and third arguments capture simulated and \invivo dependent features, respectively.

\subsection{Data Summary}

Our networks operate on time delayed aperture domain signals to perform a regression-based beamforming for each received spatial location. A Hilbert transform was applied to all received channel data prior to network processing to generate real and imaginary components. Training and test examples were generated from simulated anechoic cyst data as well as \invivo liver data. Although trivial, anechoic cysts provide clean and intuitive ground truth information and ensure an obvious domain shift between simulated and \invivo data. 

\subsubsection{Training Data}

\begin{figure}[t]
	\centering
	\includegraphics[width=3.75in]{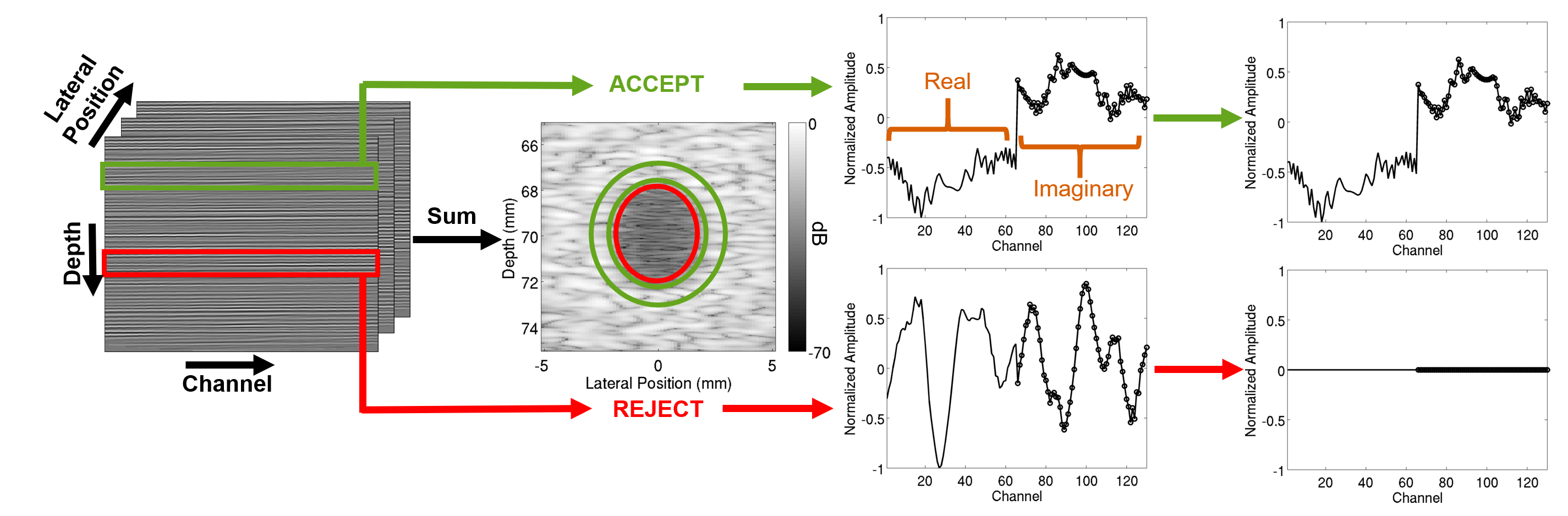}
	\caption{Example labeled simulated anechoic cyst training data. An example channel data set and corresponding B-mode image is shown on the left to indicate accept and reject regions. A total of 10 depth locations were used as input and output to each network as illustrated by the red and green boxes on the channel data. These boxes represent 10 pixels within the corresponding red and green regions in the B-mode image. Example input and output aperture domain signals are shown on the right for a single depth. Real and imaginary components are stacked to form a 1D vector as the input and output to the DNN beamformer.} \label{methodsFigDatSummary}
\end{figure}

Field II \cite{JensenF2} was used to simulate channel data of 12 5mm diameter anechoic cyst realizations focused at 70mm using a 5.208MHz center frequency, 20.832MHz sampling frequency, 1540m/s sound speed, and 65 active element channels with a pitch of 298$\mu m$. These parameters were used to mimic the probe used for acquiring the \invivo data, as described below. Simulated training data were split into accept and reject regions depending on whether the aperture signals within a 0.5$\lambda$ axial kernel (i.e., 10 depths) originated from a location outside or inside of the cyst, respectively. An output $y_s$ is taken to be the input $x_s$ if the signal is in the accept region, whereas the output $y_s$ is a vector of zeros if $x_s$ is in the reject region. Example simulated training data are shown in Fig. \ref{methodsFigDatSummary}. Each aperture domain signal was concatenated through depth in addition to concatenating real and imaginary components. The number of accept and reject training examples was made equal (i.e., the full background was not used for training). For each simulated data set, 2,782 aperture domain examples (i.e., pre-reconstructed pixels) were used during training, resulting in a total of 33,384 total paired simulated examples. 

A 36 year old healthy male gave informed written consent in accordance with the local institutional review board to acquire free-hand ultrasound channel data of his liver. A Verasonics Vantage Ultrasound System (Verasonics, Inc., Kirkland, WA) and ATL L7-4 (38mm) linear array transducer were used to acquire channel data of 15 different fields of view of the liver, 6 of which were used for training. Acquisition parameters matched those used for simulations. For each of the 6 data sets used for training, similar to what was done for the simulations, aperture domain signals originating from spatial locations within a region around the focus were extracted. The same total number of examples were used from \invivo data as were used from simulations (i.e., 33,384 unpaired \invivo examples). 

\subsubsection{Test Data} 

For testing, 21 separate anechoic cyst realizations were simulated using the same parameters as above. White gaussian noise was added to the test realizations to achieve a signal-to-noise ratio of 50dB. The remaining 9 \invivo examples not used during training were used for testing. For both the simulations and \invivo data, the full field of view was used for testing. A sliding window of 1 depth was used to select 10 depth inputs and overlapping depth outputs were averaged.

\subsection{Evaluation}

\subsubsection{Network Details} 

Network hyperparameters corresponding to layer width, number of hidden layers, and regression losses (e.g. mean squared error, $l_1$, Huber loss) were varied. Models were tested on \invivo validation data, withheld from training and testing. The model that produced the highest CNR on the validation \invivo data was selected. Additional details are included in supplementary materials.

Both convolutional \cite{hyun2019,nair2018,nair2019} and fully connected \cite{luchies2018,zhuang2019} neural networks have been investigated for the purposes of ultrasound beamforming, and it was shown previously that there was minimal difference between the two architectures \cite{chen2019}. Our networks are implicitly convolutional through depth, but fully connected across the transducer elements, which is consistent with the known signal coherence patterns of ultrasound channel data~\cite{mallart1991}. For this reason, and based on the network approach used for comparison in this work \cite{luchies2018}, all networks in this work -- generators, discriminators, and regressors -- are implemented as fully connected. 

\subsubsection{Comparison to Established Beamformers}

Both frequency \cite{luchies2018,zhuang2019} and time \cite{hyun2019,nair2018} domain approaches have been considered for ultrasound deep learning methods. Given the added complexity of our proposed training architecture and the success of other time domain implementations \cite{hyun2019,nair2018}, we use time domain data in this work. Therefore, as a direct baseline comparison to the proposed DA-DNN approach, a conventional DNN trained only on time domain simulated data, but with otherwise similar network parameters, was also evaluated. For this approach, the only loss used for optimization is summarized in Equation \ref{simLoss}. Additionally, for completeness, an established frequency-domain DNN approach \cite{luchies2018} was also evaluated. This approach differs from the aforementioned conventional DNN approach in that it uses short time Fourier transformed (STFT) data and trains separate networks for individual frequencies. For this approach, model selection was performed as in \cite{luchies2018} to highlight a best case scenario.

In addition to comparing the proposed DA-DNN approach to conventional and STFT DNN beamforming, performance was also evaluated in comparison to other established beamformers, including conventional DAS, the generalized coherence factor (GCF) \cite{li2003}, and aperture domain model image reconstruction (ADMIRE) \cite{byram2015}. For the GCF approach, as suggested by Li et al. \cite{li2003}, a cutoff spatial frequency of 3 frequency bins (i.e., $M_{0}$=1) was used to compute the weighting mask. 

\subsubsection{Performance Metrics}

Contrast-to-noise ratio (CNR) and contrast ratio (CR) were used to evaluate beamformer performance as follows,
\begin{equation}
CNR=20\log_{10}\frac{|\mu_{background}-\mu_{lesion}|}{\sqrt{\sigma^{2}_{background}+\sigma^{2}_{lesion}}} \ ; \ CR=-20\log_{10}\frac{\mu_{lesion}}{\mu_{background}}
\end{equation}
\noindent where $\mu$ and $\sigma$ are the mean and standard deviation of the uncompressed envelope. Images were made for qualitative comparison by log compressing the envelope data and scaling to a 60dB dynamic range.

\section{Results}

\begin{figure}
	\centering
	\includegraphics[width=3.75in]{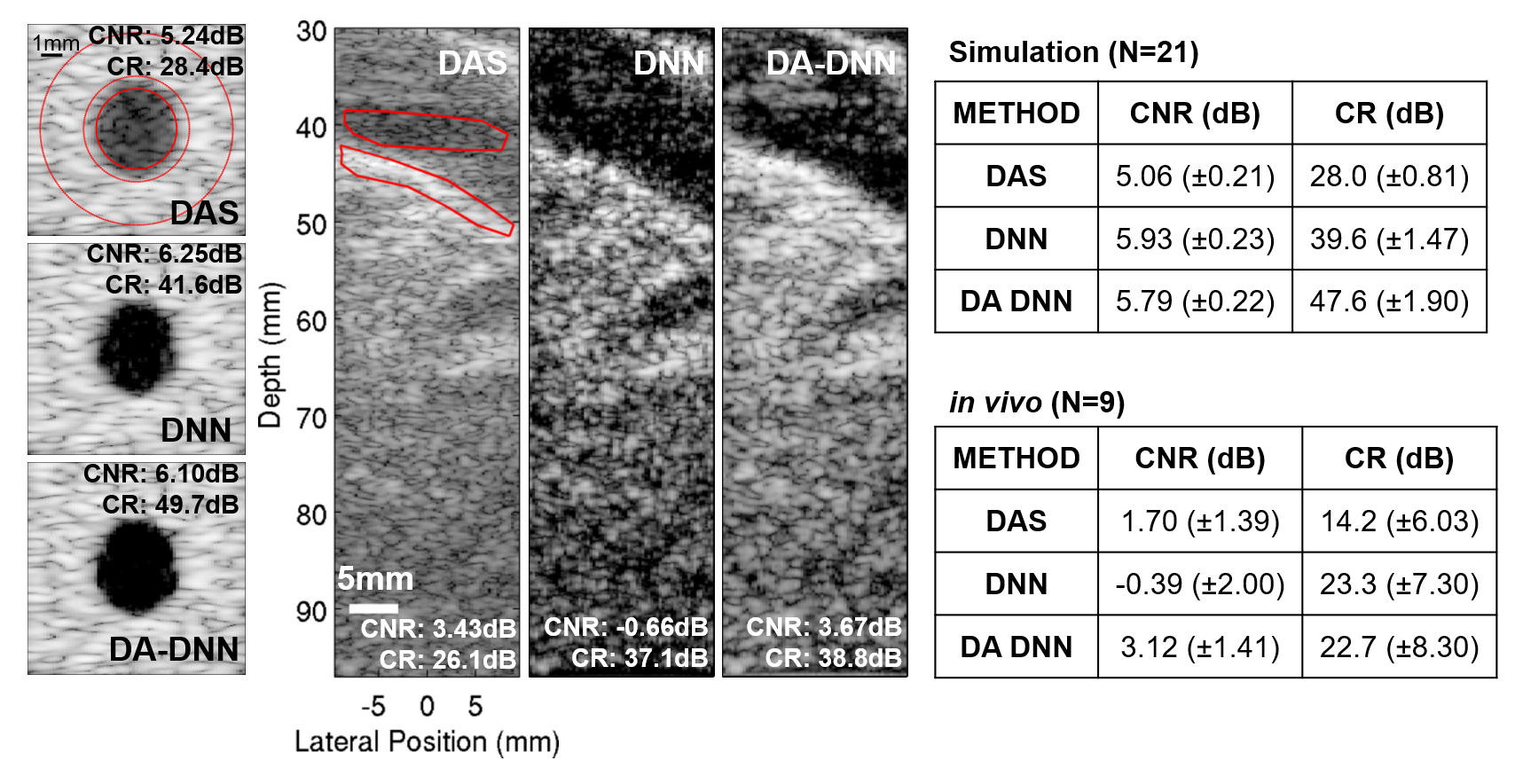}
	\caption{Example DAS, DNN, and DA-DNN results for simulated anechoic cysts and \textit{in vivo} data. All images are scaled to individual maximums and displayed with a 60dB dynamic range. Example regions of interest used to compute CNR and CR are indicated on the DAS images in red. Corresponding CNR and CR values are displayed on each image. Tables on the right indicate average CNR and CR ($\pm$ standard deviation) for each method for simulations and \invivo data.} \label{simVSinvivo}
\end{figure}

DNN beamformers trained using only simulated data work well on simulations but often fail to generalize to \textit{in vivo} data, as demonstrated in Fig. \ref{simVSinvivo}. Minimal difference was observed between the conventional DNN and DA-DNN approach when tested on simulated anechoic cysts in terms of CNR. In contrast, substantial improvements were observed both qualitatively and quantitatively when using the DA-DNN beamformer on \textit{in vivo} data. Despite producing a high CR \textit{in vivo}, the conventional DNN approach resulted in substantial drop out (i.e., extreme low amplitude pixels) in the background, resulting in a lower CNR than DAS. DA-DNN beamforming was able to maintain a higher CR compared to DAS while also improving CNR. 

\begin{figure}
	\centering
	\includegraphics[width=4in]{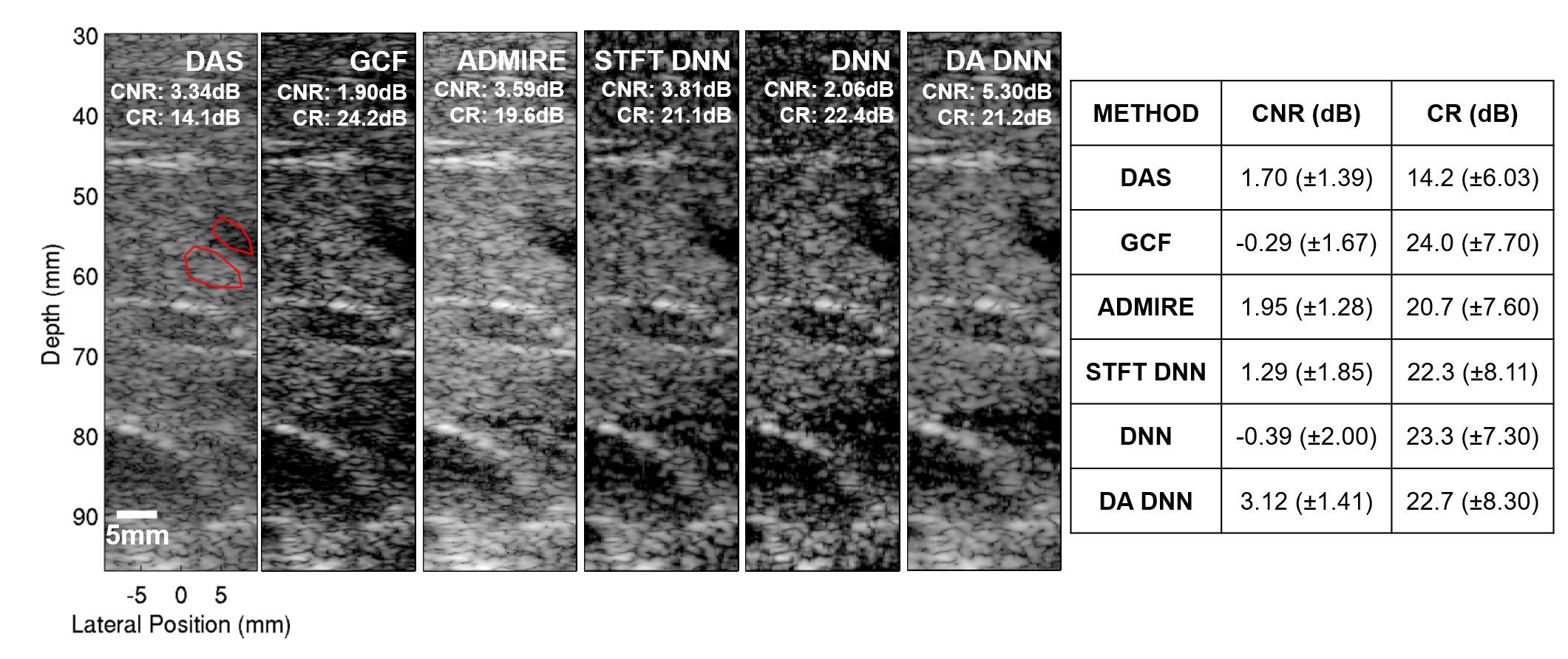}
	\caption{Example \textit{in vivo} B-mode images are shown for each beamformer. The regions of interest used to compute image quality metrics for the displayed example are shown in red on the DAS B-mode image. All images are scaled to individual maximums and a 60dB dynamic range. Average CNR and CR ($\pm$ standard deviation) across all 9 \invivo test examples are indicated in the table on the right.} \label{methodsComp}
\end{figure}

Qualitative and quantitative improvements in image quality with the DA-DNN beamformer were observed compared to the evaluated beamformers, as shown in Fig. \ref{methodsComp}. GCF and conventional DNN beamformers produce noticeably better contrast than DAS, but they also result in more drop out regions compared to ADMIRE and DA-DNN. These trends are described quantitatively by the corresponding tables, for which DA-DNN produced the highest average CNR overall while still maintaining higher CR than DAS. 

\section{Conclusion}

Conventional DNN adaptive beamforming relies on ground truth training data which is difficult to acquire \textit{in vivo}. To address this challenge, we propose a domain adaptation scheme to train an \textit{in vivo} beamformer absent of any labeled \textit{in vivo} data. We demonstrated substantial image quality improvements using our proposed approach compared to conventional DNN beamforming and to other established beamformers, including DAS, GCF, and ADMIRE. We show that DA-DNN beamforming achieved image quality consistent with or higher than state of the art ADMIRE without the same computational limitations. As stated throughout, an important fundamental assumption of our approach is that the domain shift between simulated and \invivo data is the same for the inputs and the outputs. Based on our results, this seems to be a reasonable baseline assumption, but it’s worth investigating this further in future work.

%
%
%
\bibliographystyle{splncs04}
\bibliography{paper2756_references}

\begin{thebibliography}{10}
\providecommand{\url}[1]{\texttt{#1}}
\providecommand{\urlprefix}{URL }
\providecommand{\doi}[1]{https://doi.org/#1}

\bibitem{byram2015}
Byram, B., Dei, K., Tierney, J., Dumont, D.: A model and regularization scheme
  for ultrasonic beamforming clutter reduction. IEEE transactions on
  ultrasonics, ferroelectrics, and frequency control  \textbf{62}(11),
  1913--1927 (2015)

\bibitem{byram2014}
Byram, B., Jakovljevic, M.: Ultrasonic multipath and beamforming clutter
  reduction: a chirp model approach. IEEE transactions on ultrasonics,
  ferroelectrics, and frequency control  \textbf{61}(3),  428--440 (2014)

\bibitem{chen2019}
Chen, Z., Luchies, A., Byram, B.: Compact convolutional neural networks for
  ultrasound beamforming. In: 2019 IEEE International Ultrasonics Symposium
  (IUS). pp. 560--562. IEEE (2019)

\bibitem{dahl2014}
Dahl, J.J., Sheth, N.M.: Reverberation clutter from subcutaneous tissue layers:
  Simulation and in vivo demonstrations. Ultrasound in medicine \& biology
  \textbf{40}(4),  714--726 (2014)

\bibitem{daum2007}
Daum{\'e}~III, H.: Frustratingly easy domain adaptation. In: Proceedings of ACL
  (2007)

\bibitem{dei2017}
Dei, K., Byram, B.: The impact of model-based clutter suppression on cluttered,
  aberrated wavefronts. IEEE transactions on ultrasonics, ferroelectrics, and
  frequency control  \textbf{64}(10),  1450--1464 (2017)

\bibitem{dei2018}
Dei, K., Byram, B.: A robust method for ultrasound beamforming in the presence
  of off-axis clutter and sound speed variation. Ultrasonics  \textbf{89},
  34--45 (2018)

\bibitem{dei2019}
Dei, K., Schlunk, S., Byram, B.: Computationally efficient implementation of
  aperture domain model image reconstruction. IEEE transactions on ultrasonics,
  ferroelectrics, and frequency control  \textbf{66}(10),  1546--1559 (2019)

\bibitem{gasse2017}
Gasse, M., Millioz, F., Roux, E., Garcia, D., Liebgott, H., Friboulet, D.:
  High-quality plane wave compounding using convolutional neural networks. IEEE
  transactions on ultrasonics, ferroelectrics, and frequency control
  \textbf{64}(10),  1637--1639 (2017)

\bibitem{glorot2010}
Glorot, X., Bengio, Y.: Understanding the difficulty of training deep
  feedforward neural networks. In: Proceedings of the thirteenth international
  conference on artificial intelligence and statistics. pp. 249--256 (2010)

\bibitem{relu2011}
Glorot, X., Bordes, A., Bengio, Y.: Deep sparse rectifier neural networks. In:
  Proceedings of the fourteenth international conference on artificial
  intelligence and statistics. pp. 315--323 (2011)

\bibitem{goodfellow2014}
Goodfellow, I., Pouget-Abadie, J., Mirza, M., Xu, B., Warde-Farley, D., Ozair,
  S., Courville, A., Bengio, Y.: Generative adversarial nets. In: Advances in
  neural information processing systems. pp. 2672--2680 (2014)

\bibitem{he2015}
He, K., Zhang, X., Ren, S., Sun, J.: Delving deep into rectifiers: Surpassing
  human-level performance on imagenet classification. In: Proceedings of the
  IEEE international conference on computer vision. pp. 1026--1034 (2015)

\bibitem{hoffman2017}
Hoffman, J., Tzeng, E., Park, T., Zhu, J.Y., Isola, P., Saenko, K., Efros,
  A.A., Darrell, T.: Cycada: Cycle-consistent adversarial domain adaptation.
  arXiv preprint arXiv:1711.03213  (2017)

\bibitem{holfort2009}
Holfort, I.K., Gran, F., Jensen, J.A.: Broadband minimum variance beamforming
  for ultrasound imaging. IEEE transactions on ultrasonics, ferroelectrics, and
  frequency control  \textbf{56}(2),  314--325 (2009)

\bibitem{hyun2019}
Hyun, D., Brickson, L.L., Looby, K.T., Dahl, J.J.: Beamforming and speckle
  reduction using neural networks. IEEE transactions on ultrasonics,
  ferroelectrics, and frequency control  \textbf{66}(5),  898--910 (2019)

\bibitem{isola2017}
Isola, P., Zhu, J.Y., Zhou, T., Efros, A.A.: Image-to-image translation with
  conditional adversarial networks. In: Proceedings of the IEEE conference on
  computer vision and pattern recognition. pp. 1125--1134 (2017)

\bibitem{JensenF2}
Jensen, J.A.: Field: A program for simulating ultrasound systems. Med. Biol.
  Eng. Comput.  \textbf{34},  351--353 (1996)

\bibitem{khan2019}
Khan, S., Huh, J., Ye, J.C.: Universal deep beamformer for variable rate
  ultrasound imaging. arXiv preprint arXiv:1901.01706  (2019)

\bibitem{kingma2014}
Kingma, D.P., Ba, J.: Adam: A method for stochastic optimization. arXiv
  preprint arXiv:1412.6980  (2014)

\bibitem{lediju2011}
Lediju, M.A., Trahey, G.E., Byram, B.C., Dahl, J.J.: Short-lag spatial
  coherence of backscattered echoes: Imaging characteristics. IEEE transactions
  on ultrasonics, ferroelectrics, and frequency control  \textbf{58}(7),
  1377--1388 (2011)

\bibitem{li2003}
Li, P.C., Li, M.L.: Adaptive imaging using the generalized coherence factor.
  IEEE transactions on ultrasonics, ferroelectrics, and frequency control
  \textbf{50}(2),  128--141 (2003)

\bibitem{luchies2018}
Luchies, A.C., Byram, B.C.: Deep neural networks for ultrasound beamforming.
  IEEE transactions on medical imaging  \textbf{37}(9),  2010--2021 (2018)

\bibitem{luchies2019}
Luchies, A.C., Byram, B.C.: Training improvements for ultrasound beamforming
  with deep neural networks. Physics in medicine and biology  \textbf{64}
  (2019)

\bibitem{mallart1991}
Mallart, R., Fink, M.: The van cittert--zernike theorem in pulse echo
  measurements. The Journal of the Acoustical Society of America
  \textbf{90}(5),  2718--2727 (1991)

\bibitem{mescheder2018training}
Mescheder, L., Geiger, A., Nowozin, S.: Which training methods for gans do
  actually converge? arXiv preprint arXiv:1801.04406  (2018)

\bibitem{nair2018}
Nair, A.A., Tran, T.D., Reiter, A., Bell, M.A.L.: A deep learning based
  alternative to beamforming ultrasound images. In: 2018 IEEE International
  Conference on Acoustics, Speech and Signal Processing (ICASSP). pp.
  3359--3363. IEEE (2018)

\bibitem{nair2019}
Nair, A.A., Tran, T.D., Reiter, A., Bell, M.A.L.: A generative adversarial
  neural network for beamforming ultrasound images: Invited presentation. In:
  2019 53rd Annual Conference on Information Sciences and Systems (CISS).
  pp.~1--6. IEEE (2019)

\bibitem{pytorch2017}
Paszke, A., Gross, S., Chintala, S., Chanan, G., Yang, E., DeVito, Z., Lin, Z.,
  Desmaison, A., Antiga, L., Lerer, A.: Automatic differentiation in pytorch
  (2017)

\bibitem{perdios2017}
Perdios, D., Besson, A., Arditi, M., Thiran, J.P.: A deep learning approach to
  ultrasound image recovery. In: 2017 IEEE International Ultrasonics Symposium
  (IUS). pp.~1--4. Ieee (2017)

\bibitem{senouf2018}
Senouf, O., Vedula, S., Zurakhov, G., Bronstein, A., Zibulevsky, M.,
  Michailovich, O., Adam, D., Blondheim, D.: High frame-rate cardiac ultrasound
  imaging with deep learning. In: Medical Image Computing and Computer Assisted
  Intervention -- MICCAI 2018. pp. 126--134. Springer International Publishing
  (2018)

\bibitem{simson2019}
Simson, W., G{\"o}bl, R., Paschali, M., Kr{\"o}nke, M., Scheidhauer, K., Weber,
  W., Navab, N.: End-to-end learning-based ultrasound reconstruction. arXiv
  preprint arXiv:1904.04696  (2019)

\bibitem{synnevag2007}
Synnevag, J.F., Austeng, A., Holm, S.: Adaptive beamforming applied to medical
  ultrasound imaging. IEEE transactions on ultrasonics, ferroelectrics, and
  frequency control  \textbf{54}(8),  1606--1613 (2007)

\bibitem{yoon2018}
Yoon, Y.H., Khan, S., Huh, J., Ye, J.C.: Efficient b-mode ultrasound image
  reconstruction from sub-sampled rf data using deep learning. IEEE
  transactions on medical imaging  \textbf{38}(2),  325--336 (2018)

\bibitem{zhu2017}
Zhu, J.Y., Park, T., Isola, P., Efros, A.A.: Unpaired image-to-image
  translation using cycle-consistent adversarial networks. In: Proceedings of
  the IEEE international conference on computer vision. pp. 2223--2232 (2017)

\bibitem{zhuang2019}
Zhuang, R., Chen, J.: Deep learning based minimum variance beamforming for
  ultrasound imaging. In: Smart Ultrasound Imaging and Perinatal, Preterm and
  Paediatric Image Analysis, pp. 83--91. Springer (2019)

\end{thebibliography}

\end{document}